\documentclass[preprint,12pt]{elsarticle}

\journal{European Journal of Mechanics B / Fluids}

\usepackage{amsthm}
\usepackage{amsmath,mathtools}
\usepackage{amsfonts}
\usepackage{amssymb}
\usepackage{graphicx}
\usepackage{epstopdf}
\usepackage{bm}

\begin{document}

\begin{frontmatter}
\title{Short-wave vortex instability in stratified flow}
%\author{L. Bovard\fnref{myfootnote}}
\author[Waterloo,current,contact]{L. Bovard}
\author[Waterloo]{M.L. Waite} 
\fntext[Waterloo]{Department of Applied Mathematics, University of Waterloo, 200 University Avenue West, Waterloo, Ontario N2L 3G1, Canada}
\fntext[current]{Present Address: Institute for Theoretical Physics, Frankfurt}
\fntext[contact]{Corresponding author: lbovard@uwaterloo.ca, Phone: 519.888.4567 ext. 35596}
\date{\today}
\begin{abstract}
In this paper we investigate a new instability of the Lamb-Chaplygin dipole in a stratified fluid. Through numerical linear stability analysis, a secondary peak in the growth rate emerges at vertical scales about an order of magnitude smaller than the buoyancy scale $L_{b}=U/N$ where $U$ is the characteristic velocity and $N$ is the Brunt-V\"{a}is\"{a}l\"{a} frequency. This new instability exhibits a growth rate that is similar to, and even exceeds, that of the zigzag instability, which has the characteristic length of the buoyancy scale. This instability is investigated for a wide range of Reynolds numbers, $Re=2000-20000$, and horizontal Froude numbers, $F_{h}=0.05-0.2$, where $F_{h}=U/NR$, $Re=UR/\nu$, $R$ is the radius of the dipole, and $\nu$ is the kinematic viscosity. A range of vertical scales is explored from above the  buoyancy scale to the viscous damping scale. Additionally, evidence is presented that the length scale and growth rate of this new instability are partially determined by the buoyancy Reynolds number, $Re_{b}=F_{h}^{2}Re$. 
\end{abstract}
\end{frontmatter}

%%%%%%%%%%%%%%%%%%%%%%%%%%%%%%
% INTRODUCTION
%%%%%%%%%%%%%%%%%%%%%%%%%%%%%%

\section{Introduction} 
Vortices play a fundamental role in the transition to turbulence by providing the mechanism for the energy cascade from large to small scales. In the atmosphere and ocean, vortices are strongly influenced by density stratification and the rotation of the earth. However, stratification dominates at intermediate length scales -- the atmospheric mesoscale and the oceanic submesoscale --  which are small enough for the Coriolis effects to be weak, but large enough for the stable density stratification to be strong (e.g. \cite{riley2003,waitebartello2004,rileylindborg2013}).  There has recently been much work, using full direct numerical simulations of the Boussinesq equations with various initial configurations, to uncover the emergence and evolution of stratified turbulence from vortices \cite{waitesmol2008,delonclebc2008,augierbillant2011,augier2012}. Turbulence in this regime is governed by the Reynolds number $Re=UR/\nu$ as well as the horizontal Froude number $F_{h}=U/NR$, where $U$ is the characteristic velocity, $R$ is the characteristic horizontal length, $N$ is the Brunt-V\"{a}is\"{a}l\"{a} frequency, and $\nu$ is the kinematic viscosity. Because of this extra dependence on the Froude number, the underlying dynamics are not as well understood and a full picture of stratified turbulence is not complete \cite{rileylelong2000,riley2003,waitebartello2004,lindborg2006,rileylindborg2013}.  

In large-scale atmosphere and ocean simulations, it is difficult or impossible to resolve all possible processes. As a result, obtaining a proper parameterisation of small-scale phenomena is critical to correctly modelling the evolution. A useful approach to investigating these small-scale dynamics is to consider the transition problem in an idealised flow, which can elucidate the key features that govern the more comprehensive turbulence problem. One model that may be used to study the transition to stratified turbulence is that of a columnar counter-rotating vortex dipole. There is a large body of literature on the instability of vortex dipoles in unstratified fluids, including the Crow instability at large length scales (e.g. \cite{crow1970,widnall1974,leweke1998b}) and the elliptic instability at smaller scales (e.g. \cite{widnall1974,pierrehumbert1986,baily1986,waleffe1990}). In stratified fluids, laboratory and numerical experiments of the stability of such dipoles have uncovered a unique instability, the zigzag instability, so named due to the zigzag-like structure exhibited by the flow \cite{bc2000a,bc2000c}. The zigzag instability has a dominant vertical wavelength of around $U/N$, which is known as the buoyancy scale \cite{waite2011}. This instability has also been found in other flow configurations including co-rotating vortices \cite{otheguybc} and vortex arrays \cite{delonclebc2011}. The breakdown of this dipole into turbulence due to the growth and saturation of the zigzag instability has also been investigated \cite{waitesmol2008,augierbillant2011,delonclebc2008}. However, these studies mainly consider dipoles perturbed at the zigzag wavelength $U/N$, and do not investigate the growth of smaller vertical scale perturbations. Growth in such small-scale perturbations has been reported in nonlinear simulations \cite{waitesmol2008}. In this work we investigate the linear stability of the dipole at these small vertical scales. 

The buoyancy scale is an important length scale in stratified turbulence. It is the vertical scale at which the vertical Froude number is $O(1)$ \cite{bc2001}, and it naturally emerges as the thickness of layers in stratified turbulence \cite{bc2001,waitebartello2004}. There is a direct transfer of energy, believed to be due to Kelvin-Helmholtz instability \cite{waite2011,augier2012}, from large horizontal scales into the buoyancy scale in stratified turbulence \cite{waite2011} and in the breakdown of the zigzag instability \cite{augier2012}. This breakdown generates small-scale turbulence which ultimately fills the spectrum at scales below the buoyancy scale. But it is possible that primary instabilities of the large-scale vortex may also directly excite vertical scales below the buoyancy scale. We investigate this possibility here.  

In this paper we extend the linear stability analysis of the zigzag instability \cite{bc2000c} by investigating short, sub-buoyancy scale vertical wavelength perturbations of the Lamb-Chaplygin dipole in a stratified flow. The Lamb-Chaplygin dipole, an exact 2D solution to the Euler equations, is a good approximation to columnar counter-rotating dipole generated in lab experiments \cite{bc2000a}.  The work is presented as follows: in section 2 we present the numerical scheme and methodology, in section 3 we discuss the results of the numerical simulations and investigate some properties of the small-scale instability. Conclusions are discussed in the last section. 

%old intro comments 
%In their study Waite \& Smolarkiewicz\cite{waitesmol2008} noticed instability at short vertical scales of the columnar counter-rotating dipole and in this paper we investigate it further. Pierrehumbert \cite{pierrehumbert1986} initially looked at inviscid unstratified vortices subject to three dimensional perturbations and presented evidence that the growth rate is independent of wave number for small vertical scales. Miyazaki \& Fukumoto\cite{miyazakifukumoto1992}, motivated by \cite{pierrehumbert1986}, investigated the same problem but with stratification and found that the results of \cite{pierrehumbert1986} were suppressed above a critical Brunt-V\"{a}is\"{a}l\"{a} frequency. 
%By having a complete picture of the transition to stratified turbulence, improvements can be made to the full nonlinear simulations as potentially important mechanisms can be correctly resolved \cite{waitesmol2008,augierbillant2011,delonclebc2008}. 
%Understanding how stratified turbulence differs from unstratified turbulence is important for large scale turbulence simulations to ensure that all relevant length scales and phenomena are captured in the numerical simulations \cite{waitesmol2008,augierbillant2011,delonclebc2008,waite2011}. A common approach to studying turbulence is to investigate the transition problem in simplified models that can illustrate the key features and behaviour that govern the more complicated turbulent flow \cite{waitesmol2008,augierbillant2011,delonclebc2008}. 

%%%%%%%%%%%%%%%%%%%%%%%%%%%%%%
% FORMULATION
%%%%%%%%%%%%%%%%%%%%%%%%%%%%%%

\section{Formulation}
\subsection{Equations and Initial Conditions}
We consider the non-dimensional Boussinesq approximation to the Navier-Stokes equations in Cartesian co-ordinates
\begin{align}
\frac{D\bm{u}}{Dt} = -\nabla p - \rho'\hat{\bm{e}}_{z} + \frac{1}{Re}\nabla^{2} \bm{u},\\
\nabla \cdot \bm{u}=0,\\
\frac{D\rho'}{Dt} -\frac{w}{F_{h}^{2}} = \frac{1}{ReSc}\nabla^{2} \rho',
\end{align}
where $D/Dt=\partial/\partial t + \bm{u}\cdot \nabla, \bm{u}=(u,v,w)$ is the velocity, $p$ is the pressure, and $\rho'$ is the density perturbation. We have non-dimensionalised by the characteristic velocity $U$, length $R$, time-scale $R/U$, pressure $\rho_{0}U^{2}$, density $\rho_{0}U^{2}/gR$, and defined $Sc=\nu /D$ as the Schmidt number, where $D$ is the mass diffusivity, $\rho_{0}$ is the background density, and $g$ is the gravitational acceleration. The Reynolds and horizontal Froude number are defined as above. The buoyancy frequency $N$, and hence the Froude number $F_{h}$, is assumed to be constant. 

As the basic state for linear stability analysis we use the Lamb-Chaplygin dipole in a comoving frame \cite{meleshko1994}. This dipole is a solution to the 2D inviscid Euler equations. This basic state is motivated by laboratory experiments \cite{bc2000a,leweke1998} which demonstrated that a vertically oriented Lamb-Chaplygin dipole is a good approximation to the vortex generated by two flaps closing in a tank of salt-stratified water. The dipole, in cylindrical coordinates, is given by the stream function
\begin{align}
\psi_{0}(r,\theta) = 
\begin{dcases}
-\frac{2}{\mu_{1}J_{0}(\mu_{1})}J_{1}(\mu_{1}r)\sin\theta & r\le 1,\\
-r\left(1-\frac{1}{r^{2}}\right)\sin\theta & r>1 ,
\end{dcases}
\end{align}
and the corresponding vertical vorticity $\omega_{z0}=\nabla_{h}^{2}\psi_{0}$
\begin{align}
\omega_{z0}(r,\theta) = \
\begin{dcases}
\mu_{1}^{2}\psi_{0}(r,\theta) & r\le 1,\\
0 & r>1,
\end{dcases}
\end{align}
where $J_{0},J_{1}$ are the zero and first order Bessel functions, $\mu_{1}\approx 3.38317$ is the first root of $J_{1}$, and $\nabla_{h}$ is the horizontal Laplacian. The basic state velocity is purely horizontal and is given by $\bm{u}_{h0}=\nabla_{h}\psi_{0}\times\hat{\bm{e}}_{z}$.

We now write the fields as the basic state plus perturbations, denoted by $\sim$. Ignoring the viscous diffusion of the basic state \cite{drazinreid} and neglecting products of the perturbations, we obtain the following set of linear equations for the perturbations
\begin{align}
\frac{\partial \tilde{\bm{u}}}{\partial t} + \omega_{z0}\hat{\bm{e}}_{z}\times \tilde{\bm{u}}+\tilde{\boldsymbol{\omega}}\times \bm{u}_{h0} = -\nabla(\tilde{p}+\bm{u}_{h0} \cdot \tilde{\bm{u}}) - \tilde{\rho}'\hat{\bm{e}}_{z} + \frac{1}{Re}\nabla^{2}\tilde{\bm{u}},\label{nsl1}\\
\nabla\cdot\tilde{\bm{u}}=0,\\
\frac{\partial \tilde{\rho}'}{\partial t} + \bm{u}_{h0}\cdot \nabla_{h}\tilde{\rho}'-\frac{1}{F_{h}^{2}}\tilde{w} = \frac{1}{ScRe}\nabla^{2}\tilde{\rho}',\label{nsl3}
\end{align}
where $\tilde{\boldsymbol{\omega}}=\nabla \times \tilde{\bm{u}}$.

As stated above, the Lamb-Chaplygin dipole is oriented vertically. As a result we can separate the perturbation into the vertical and horizontal directions as 
\begin{align} 
[\tilde{\bm{u}},\tilde{p},\tilde{\rho}'](x,y,z,t) = [\bm{u},p,\rho'](x,y,t)e^{ik_{z}z} + \text{c.c.},
\end{align}
where c.c. is the complex conjugate. From here we can now take the 2D Fourier transform and define a projection operator $\textbf{P}(\textbf{k})$, with components $P_{ij}(\textbf{k})=\delta_{ij} - k_{i}k_{j}/k^{2}$ to eliminate pressure (e.g. Lesieur \cite{lesieur2008turbulence}) to obtain a set of equations for the Fourier coefficients 

\begin{align}
\frac{\partial \hat{\bm{u}}}{\partial t} = \textbf{P}(\textbf{k})[\widehat{\bm{u}\times \omega_{z0}\hat{\bm{e}}_{z}} + \widehat{\bm{u}_{h0}\times\bm{\omega}}-\hat{\rho}'\hat{\bm{e}}_{z}] - \frac{k^{2}}{Re}\hat{\bm{u}},\label{solve1}\\
\frac{\partial\hat{\rho}'}{\partial t} = -i\bm{k}_{h}\cdot\widehat{\bm{u}_{h0}\rho'} + \frac{1}{F_{h}^{2}}\hat{w}- \frac{k^{2}}{ScRe}\hat{\rho}',\label{solve2}
\end{align}
where $k_{z},Re,Sc,F_{h}$ are input parameters, $\bm{k}_{h}=(k_{x},k_{y})$ is the horizontal wavenumber,  $k^{2}=k_{x}^{2}+k_{y}^{2}+k_{z}^{2}$ is the total wavenumber, and hat denotes Fourier coefficients. 

\subsection{Numerical Scheme}

To numerically solve (\ref{solve1}) and (\ref{solve2}), we use a spectral transform method to evaluate derivatives, with 2/3-rule de-aliasing and second order Adams-Bashforth for time-stepping. The viscous and diffusive terms are treated exactly with an integrating factor technique. Each simulation was initialised with a random density and velocity field and integrated over an $N\times N$ grid for 100 time units to determine the behaviour of the fastest growing mode. After several time units, the leading eigenmodes for $\bm{u},\rho$ behave exponentially 
\begin{align}
\bm{u},\rho \propto C(x,y)e^{\sigma t},
\end{align}
and we can obtain the largest growth rate by the formula
\begin{align}
\sigma = \lim_{t\rightarrow\infty}\frac{1}{2}\frac{d\ln E}{dt}\label{sigma1},
\end{align}
where $\sigma$ is the real growth rate of the mode and $E$ is the kinetic energy $\frac{1}{2}(u^{2}+v^{2}+w^{2})$ (e.g. Billant and Chomaz \cite{bc2000c}). To evaluate $\sigma$, we compute the average value of the growth rate beginning at $t=20$, after the initial transient behaviour has died out and the leading mode dominants, from the time series of $\sigma$  produced by (\ref{sigma1}) until the end time $t=100$. In the case of an oscillatory growth rate, as considered in \cite{bc1999}, we drop the assumption that $\sigma$ is real and instead compute the growth rate from
\begin{align}
\sigma_{r} = \lim_{t\rightarrow \infty} \frac{1}{2T}\ln\left(\frac{E(t+T)}{E(t)}\right)\label{sigma2},
\end{align}
where $T$ is the period of the oscillatory mode. The imaginary growth rate is given as $\sigma_{i}=2\pi/T$. As above, we compute $\sigma_{r}$ from the time series beginning at $t=20$, however we first measure the period $T$ from roughly 10 oscillations, and then compute the average.  

For our simulations a grid size of $L=9$ with $N=512$ points was used with timestep $\Delta t=0.000950$ for $F_{h}=0.2,Re=2000,5000,10000$ and $\Delta t=0.000375$ for all the other simulations. Unlike Billant and Chomaz \cite{bc2000c} we did not restart each simulation with the previous eigenmode because we used a parallel approach for evaluating multiple $k_{z}$ simultaneously. We investigate a range of Froude and Reynolds numbers and a wide range of $k_{z}$ from $1$ to $200$ depending on the Froude and Reynolds number. This wavenumber range incorporates the scale of the zigzag instability down to the viscous damping scale. We take $Sc=1$ for all simulations.  

To investigate even higher Reynolds numbers, we use a hyperviscosity operator. The $\nu\nabla^{2}$ diffusion term is replaced with a $\nu_{4}\nabla^{4}$ diffusion term. The $\nu_{4}$ coefficient is chosen so that $\nu k_{max}^{2} = \nu_{4}k_{max}^{4}$, where $k$ is the maximum dealiased horizontal wave number. This allows us to define the hyperviscosity Reynolds number $Re_{h}=Re k_{max}^{2}$. The hyperviscosity simulation was run with $F_{h}=0.1$ and $Re=20000$ with the same numerical parameters as the regular viscosity simulation.

%%%%%%%%%%%%%%%%%%%%%%%%%%%%%%
% RESULTS 
%%%%%%%%%%%%%%%%%%%%%%%%%%%%%%

\section{Results}
\subsection{Growth Rate}
\begin{figure}
\begin{center}
\includegraphics[width=\textwidth]{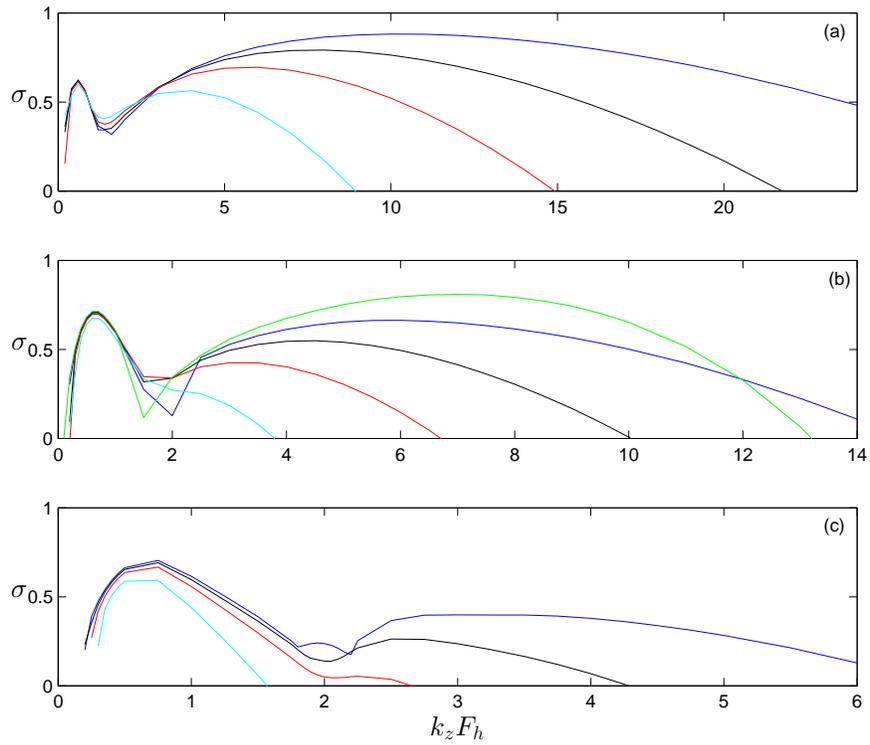}
\caption{Growth rate $\sigma$ as a function of $k_{z}F_{h}$ for fixed $F_{h}=$(a) $0.2$, (b) $0.1$, (c) $0.05$ with Re$=2000$ (cyan), Re$=5000$ (red), Re$=10000$ (black), Re$=20000$ (blue). In panel (b) the green line is the hyperviscosity case with $Re=20000$.}
\label{FixFhVaryRe}
\end{center}
\end{figure}

Fig.~\ref{FixFhVaryRe} shows the largest eigenmode growth rate as a function of vertical wavenumber for fixed $F_{h}$ and $Re$. Following Billant and Chomaz \cite{bc2000c}, the scaled vertical wavenumber $k_{z}F_{h}$ is employed. The qualitative behaviour for the growth rates at different Reynolds numbers are very similar to one another. At small $k_{z}$, the growth rate reaches a local maximum, the zigzag peak, located at $k_{z}F_{h}\approx 0.6$ as predicted by Billant and Chomaz \cite{bc2000c}.  The growth rate then decreases for increasing $k_{z}$ to a local minimum before increasing to a second local maximum. Continuing to even smaller vertical scales, viscous effects increase and may damp out the instability, and hence the growth rate decays with increasing $k_{z}F_{h}$ in the limit of large $k_{z}F_{h}$. Oscillatory growth rates are observed for the smallest $k_{z}F_{h}$ as observed in Ref 26\nocite{bc1999}. The imaginary part of the growth rate $\sigma_{i}$ remains zero everywhere else except in a small region surrounding the local minimum between the zigzag and short-wave peaks. This oscillatory behaviour is not considered here. 

\begin{figure}
\begin{center}
\includegraphics[width=\textwidth]{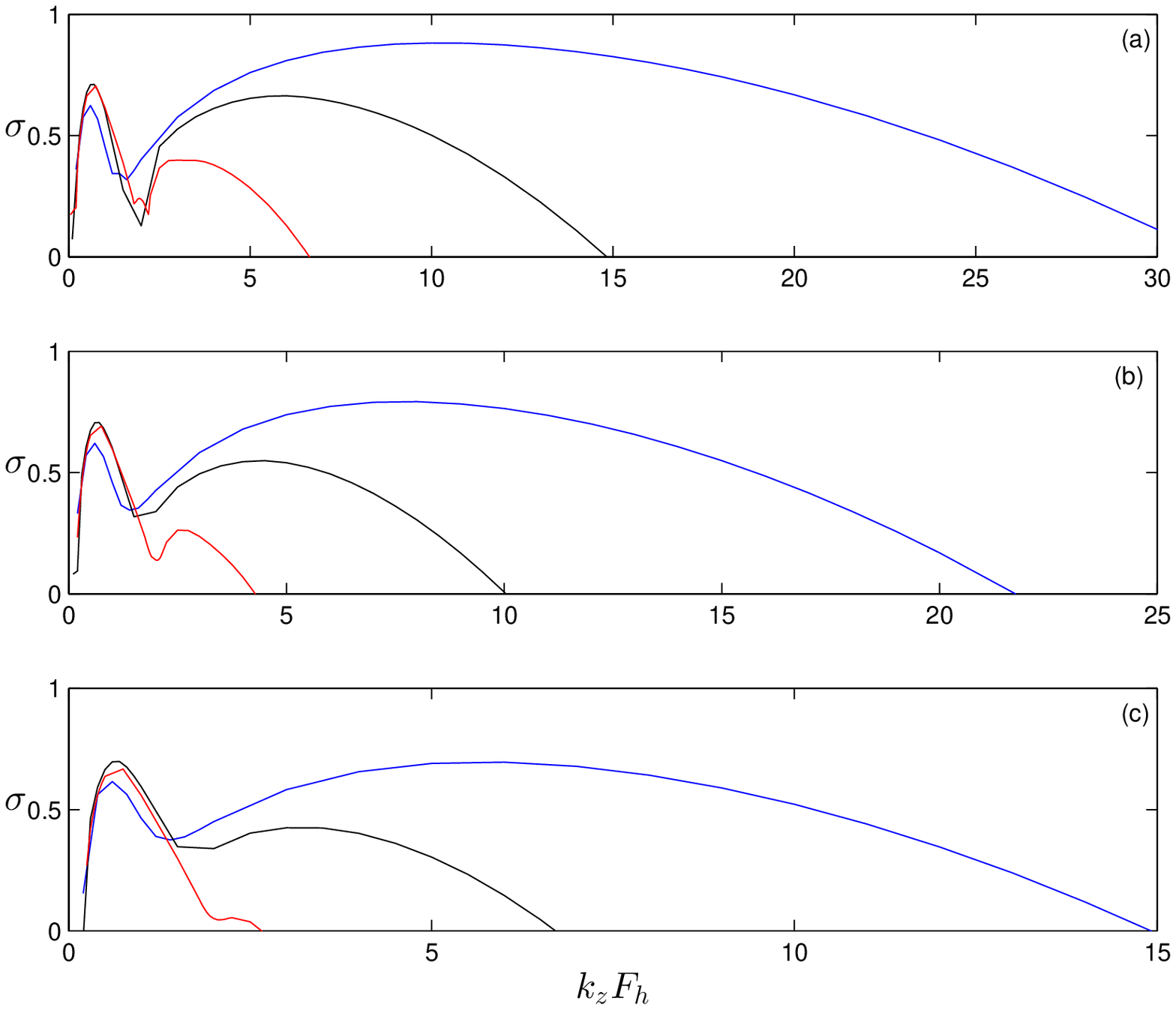}
\caption{Growth rate $\sigma$ as a function of $k_{z}F_{h}$ for fixed $\text{Re}=(a) 20000, (b) 10000, (c) 5000$ with $F_{h}=0.05$ (red), $F_{h}=0.1$ (black), $F_{h}=0.2$ (blue).}
\label{FixReVaryFh}
\end{center}
\end{figure}

For $F_{h}=0.2$ (Fig.~\ref{FixFhVaryRe}a), the peak growth rate of the short-wave instability exceeds that of the zigzag instability for increasing Reynolds numbers. The growth rates at the second peak is smaller for $F_{h}=0.1$ (Fig.~\ref{FixFhVaryRe}b), but they continue to increase with increasing $Re$. For $F_{h}=0.05$ (Fig.~\ref{FixFhVaryRe}c), the second peak is weaker than the zigzag peak. Fig.~\ref{FixReVaryFh} shows the growth rate for fixed Reynolds numbers with varying Froude numbers. Examining the case of $Re=20000$ (Fig.~\ref{FixReVaryFh}a), the second peak increases with increasing Froude. A similar result is observed for $Re=10000$ and $5000$ (Fig.~\ref{FixReVaryFh}b-c). $Re=2000$ is not included because viscous effects have damped out the second peak in this case. Overall, the dependence of the short-wave growth rate on Froude is also more pronounced then that of Reynolds. For example, the growth rate of the second peak at fixed $Re=20000$ (Fig.~\ref{FixReVaryFh}a) doubles from $F_{h}=0.05$ to $F_{h}=0.2$. By contrast, at fixed $F_{h}=0.2$ (Fig.~\ref{FixFhVaryRe}a), the increase in the growth rate from $Re=5000$ to $Re=20000$ is only about $25\%$ larger. 

\begin{figure}
\begin{center}
\includegraphics[scale=0.6]{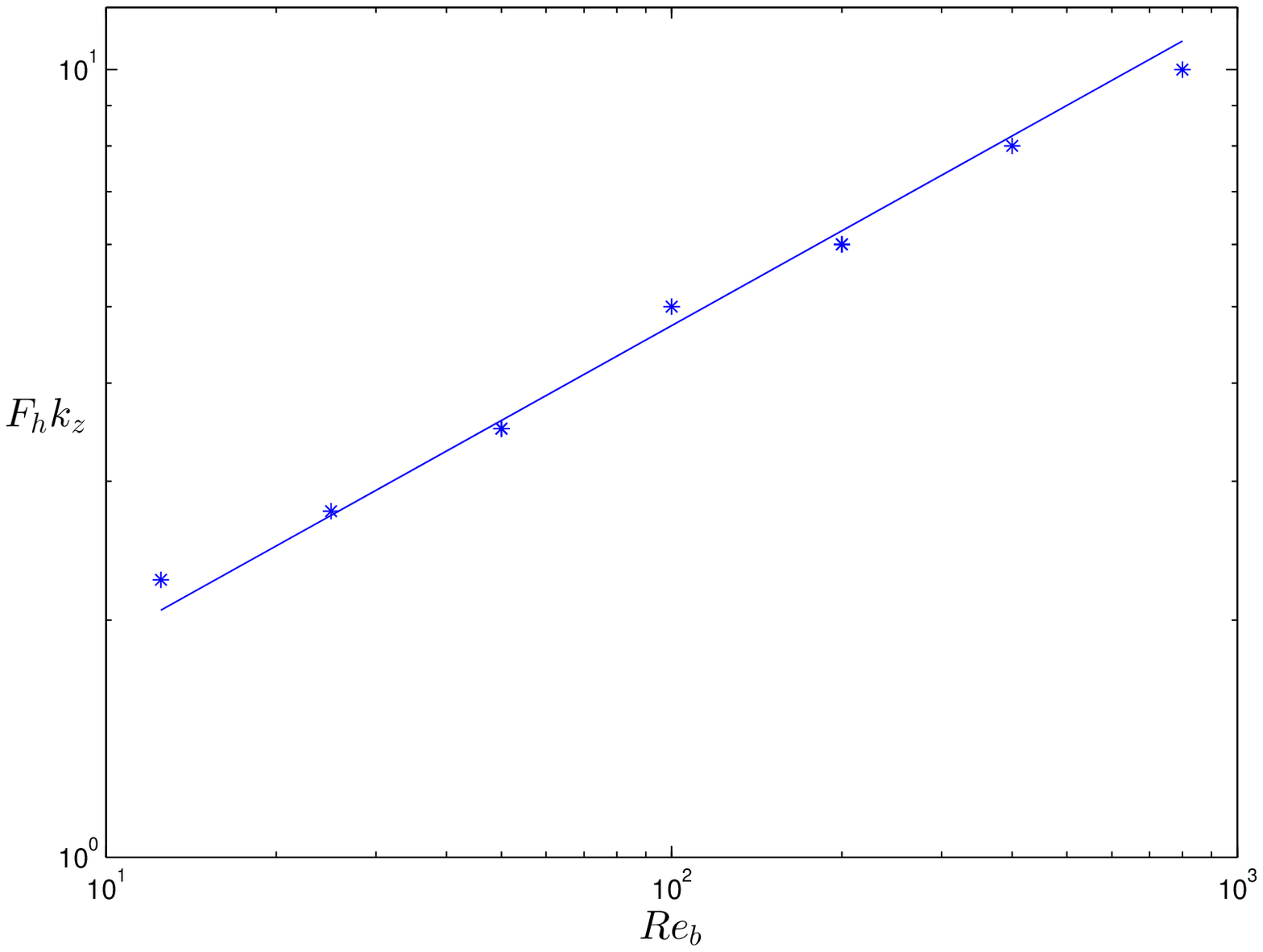}
\caption{The location of the second peak as a function of the buoyancy Reynolds number $Re_{b}$. $k_{z}F_{h}$ is taken from Fig.~\ref{FixFhVaryRe}. The straight line is $Re_{b}^{2/5}$.}
\label{Buoy}
\end{center}
\end{figure}

\begin{figure}
\begin{center}
\includegraphics[width=\textwidth]{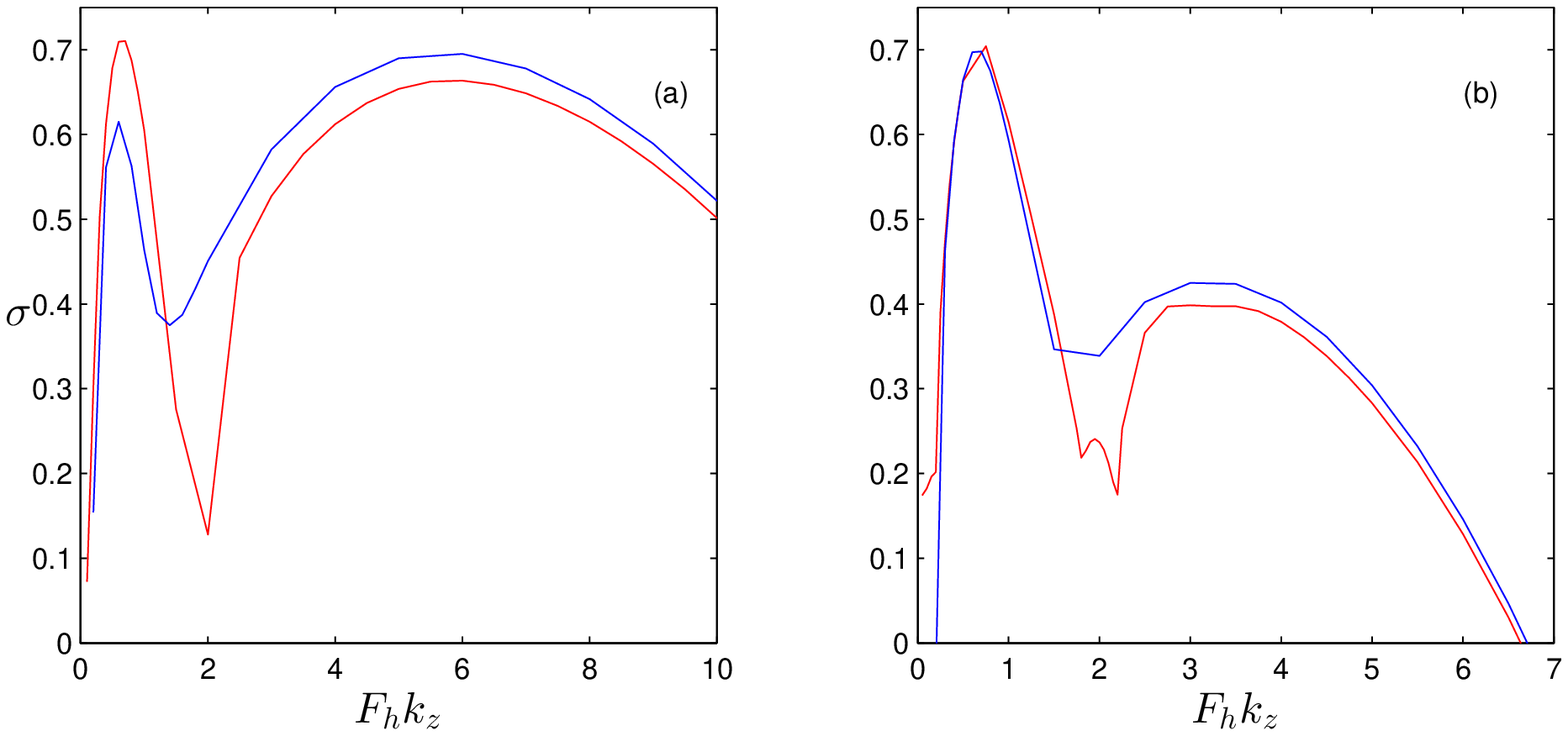}
\caption{Growth rate $\sigma$ as a function of $F_{h}k_{z}$ for fixed $Re_{b}$. In (a), red is $Re=20000, F_{h}=0.1$ and blue is $Re=5000, F_{h}=0.2$, both corresponding to $Re_{b}=500$; in (b) red is $Re=20000, F_{h}=0.05$ and blue is $Re=5000, F_{h}=0.1$, both correpsonding to $Re_{b}=50$.}
%\caption{Growth rate $\sigma$ as a function of $k_{z}$ for fixed $Re_{b}=(a) 200, (b) 50$. In (a) red corresponds to $Re=20000, F_{h}=0.1$ blue to $Re=5000, F_{h}=0.2$, in (b) red corresponds to $Re=20000, F_{h}=0.05$, blue $Re=5000, F_{h}=0.1$}
\label{ReBuoy}
\end{center}
\end{figure}

The above analysis demonstrates that the short-wave growth-rate peak moves to larger $k_{z}F_{h}$ with increasing $F_{h}$ and increasing $Re$, but has a stronger dependence on Froude than Reynolds. Some of this joint dependence can be explained by examining the dependence on the buoyancy Reynolds number $Re_{b}=F_{h}^{2}Re$ \cite{riley2003,hebert2006,brethouwer2007}. In stratified turbulence, the buoyancy Reynolds number is analogous to the Reynolds number in the viscous term due to the vertical gradients \cite{brethouwer2007}.  In Fig.~\ref{Buoy} the location of the second peak from Fig.~\ref{FixFhVaryRe} is plotted as a function of the buoyancy Reynolds number. The peak location line is approximately linear and can be fitted with the curve $k_{z}F_{h}= Re_{b}^{2/5}$, which is plotted. This scaling implies that the vertical wavenumber, $k_{z}$, of the short-wave instability is approximately 
%As $k_{z}$ increases, we move to smaller vertical scales where the vertical viscosity terms, controlled by the buoyancy Reynolds number, dominates, so it follows that the second peak may be governed by $Re_{b}$. 
\begin{align}
k_{z} \sim F_{h}^{-1/5} Re^{2/5}\label{buoyscale}.
\end{align} 
The dependence of the growth rate on $k_{z}F_{h}$ appears to be similar in the cases with different $F_{h}$ and $Re$ but the same $Re_{b}$. Fig.~\ref{ReBuoy} demonstrates the similarity of the growth rate plotted against $k_{z}F_{h}$ for two cases with $Re_{b}=500$ and two cases with $Re_{b}=50$. For both cases, the locations of the zigzag and second peak line up quite well. The difference between the red and blue curves at the second peak is $4\%$ for $Re_{b}=200$ and $6\%$ for $Re_{b}=50$, a reasonable variation. 

% It is interesting to note that the red curve, corresponding to $Re=20000$ and $F_{h}=0.1$ (a), $F_{h}=0.05$ (b), is lower then the blue curve, corresponding to $Re=5000$ and $F_{h}=0.2$ (a), $F_{h}=0.1$ (b). This supports the observation, and is clear from the definition of the buoyancy Reynolds number, that the stratification may play a more important role in the instability than the viscosity.   


In Fig.~\ref{FixFhVaryRe} (b) the green curve corresponds to a hyperviscosity run with $Re=20000$, which has $Re_{h}=2.8\times 10^{8}$. The motivation for using hyperviscosity is to capture higher-Reynolds number regime by restricting dissipation to only the largest wavenumbers. As the hyperviscosity run demonstrates, the zigzag peak is independent of Reynolds number and the existence of the peak would be expected at higher Reynolds numbers. For the second peak, we note that the growth rate  of the hyperviscosity run exceeds that of $Re=20000$ for $k_{z}F_{h}>3$ and reaches a maximum around $k_{z}F_{h}=7$. The maximum growth rate in the hyperviscosity case is around $25\%$ larger than the regular viscosity case with $Re=20000$. At $k_{z}F_{h}=12$ we see the hyperviscosity and non-hyperviscosity curves cross. This intersection corresponds to the horizontal wavenumber at which the hyperviscosity damping rate equals the regular viscous damping rate for $Re=20000$. For $k_{z}$ greater than this maximum, the hyperviscosity operator experiences greater damping than the regular viscosity, which can be seen by the sudden drop off of the growth rate. This simulation presents evidence that as $Re\rightarrow \infty$, the growth rate of the second peak will the same order as, or larger than, the growth rate of the zigzag instability. 

\subsection{Structure} 
\begin{figure}
\begin{center}
\includegraphics[width=\textwidth]{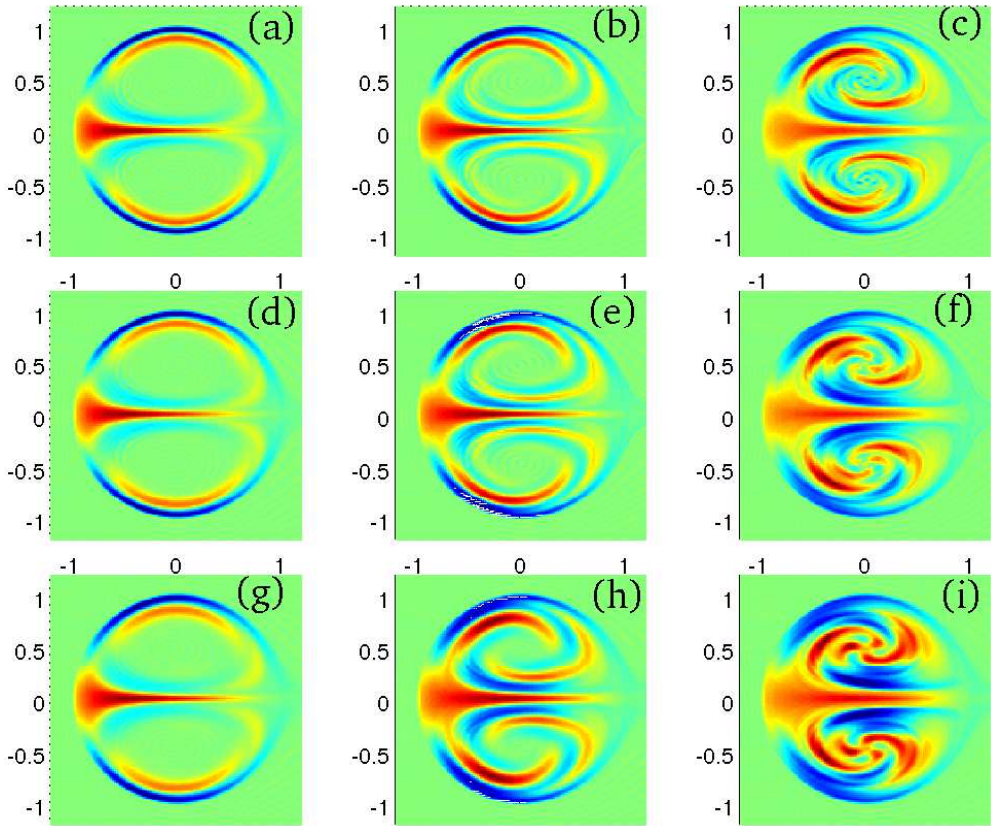}
%\caption{Perturbation vertical vorticity $\omega_{z}$ for $k_{z}$ at the second peak. $F_{h}=(a)(d)(g) 0.2 ,(b)(e)(h) 0.1, (c)(f)(i), 0.05, Re=(a)(b)(c) 20000, (d)(e)(f) 10000, (g)(h)(i) 5000$.}
\caption{Perturbation vertical vorticity $\omega_{z}$ at second peak for $Re=20000\text{ (top) }, 10000 \text{ (middle) }, 5000 \text{ (bottom) }$; and $F_{h}=0.2 \text{ (left) }, 0.1 \text{ (middle) }, 0.05 \text{ (right) }$.}
\label{secondpeak}
\end{center}
\end{figure}
Fig.~\ref{secondpeak} shows the spatial structure of the perturbation vertical vorticity at the second peak for different $Re$ and $F_{h}$. Qualitatively, we observe greater variation for different Froude numbers versus different Reynolds number as suggested above.  At the largest Froude number, the perturbation vorticity is organised in thin strips around and inside the dipole core between the two vortices. Panels (b),(e),(h) have $F_{h}=0.1$ and have a similar overall structure to the larger Froude number. Here, in the cores of the vortices, there is an emergence of a swirl-like pattern. At lower Reynolds number, the structure is spread out due to diffusion, while at higher Reynolds number, small-scale structure is beginning to emerge. This trend continues overall as we move to lower Froude numbers. 

Examining panels (g)-(i) (fixed $Re$ and decreasing $F_{h}$), the core of the dipoles has a twisting-like behaviour as the Froude number decreases. From this we can conclude that the instability structure of the second peak depends more on the Froude number than on the Reynolds number, which again reinforces the buoyancy Reynolds number scaling.  Indeed, if we consider the cases with $Re_{b}=50$ and $200$ as above, which correspond to Fig.~\ref{secondpeak} (b),(g) and (c),(h) respectively, we can see similar structure in the vorticity fields. Additionally, the anti-symmetric structure of the perturbation can be observed in the dominant eigenmodes in all cases, as found by Refs 17,26\nocite{bc1999,bc2000c}.

% The vorticity being very thin in the centre is consistent with the results of \cite{pierrehumbert1986} which examined unstratified inviscid vortices at small vertical scales which also demonstrated this behaviour. 
\begin{figure}
\begin{center}
\includegraphics[scale=0.5]{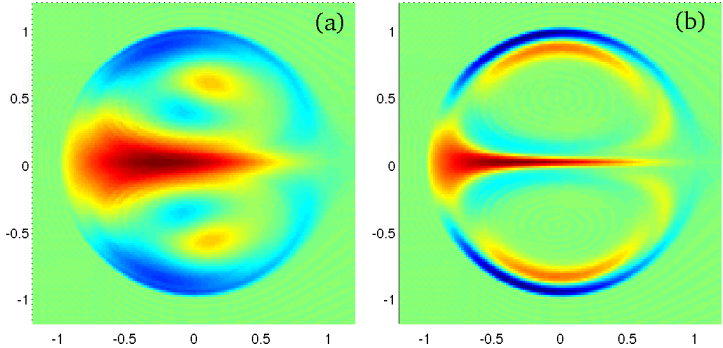}
\caption{Perturbed vertical vorticity $\omega_{z}$ at (a) the zigzag peak (b) the second peak for $Re=5000, F_{h}=0.2$}
\label{zigzagcomparison}
\end{center}
\end{figure}

Fig.~\ref{zigzagcomparison} shows the perturbation structure for the zigzag peak (a) and the short-wave peak (b) for the case of $Re=5000,F_{h}=0.2$. This case was chosen because the growth rates of the two wavenumbers is roughly the same (see Fig~\ref{FixFhVaryRe} a). The zigzag instability exhibits a quadropole vorticity structure as discussed in Ref 17\nocite{bc2000c}, which corresponds to a bend and a twist of the basic state dipole. The short-wave instability shares some common overall structure with the zigzag instability. Both have a line of vorticity centred in between two Lamb-Chaplygin vortices and have a ring of negative vorticity around the outer edges of the dipoles. Additionally, the number of local maximum and minimum remains the same. However, in the short-wave instability, these bands of vorticity have been squeezed into thinner strips and are much more localised along the outer edges of the vortices. In the cores of the dipoles, there is almost no structure and we do not see a quadropole moment. The full vorticity field of the short-wave instability has a much more dominant twist then the zigzag instability and the bending of the dipole is reduced. As the stratification is increased, this behaviour continues but there is a significant emergence of structure within the cores of the vortices, as observed in Fig~\ref{secondpeak}.

\subsection{Scale Analysis}
Motivated by the scale analysis of Refs 8,21,28,29\nocite{lilly1983,rileylelong2000,bc2001,brethouwer2007}, we present a scaling analysis for small vertical scales as considered in the above numerical simulations. We consider the Boussinesq equations

\begin{align}
\frac{\partial \textbf{u}_{h}'}{\partial t'} + \textbf{u}_{h}'\cdot\nabla'_{h}\textbf{u}_{h}'+u_{z}'\frac{\partial \textbf{u}_{h}'}{\partial z'} &= -\frac{1}{\rho_{0}}\nabla'_{h}p' + \nu \nabla'^{2}\textbf{u}_{h}\label{scaling_horz},\\
\frac{\partial u_{z}'}{\partial t'} + \textbf{u}_{h}'\cdot\nabla'_{h}u_{z}'+u_{z}'\frac{\partial u_{z}'}{\partial z'} &= -\frac{1}{\rho_{0}}\frac{\partial p'}{\partial z'} - \frac{\rho' g}{\rho_{0}} + \nu \nabla'^{2}u_{z},\label{scaling_vert}\\
\nabla'_{h}\cdot\textbf{u}_{h}' + \frac{\partial u_{z}'}{\partial z'} &=0,\label{scaling_cont}\\
\frac{\partial \rho'}{\partial t'} + \textbf{u}_{h}'\cdot\nabla_{h}'\rho' + u_{z}'\frac{\partial \rho'}{\partial z'} + \frac{\partial \rho}{\partial z'}u_{z}'&=D\nabla^{2}\rho ',\label{scaling_moment}
\end{align}
where the primed notation denotes the dimensional variables in this section only. 

Following Ref 21\nocite{bc2001} let $U,W$ be the characteristic velocities in the horizontal and vertical directions, $L_{h},L_{v}$ be the corresponding characteristic length scales, $P$ be the pressure, and $R$ be density perturbation scales, not to be confused with the dipole radius $R$ from above. We assume, differing from the analysis of Refs 21,29\nocite{lilly1983,bc2001}, that in addition to $U,L_{h}$ being imposed on the system, we also impose a separate vertical scale $L_{v}$. This scaling is motivated by the above numerical simulations where we impose a vertical length scale through the vertical wavenumber $k_{z}$. The aspect ratio $\delta=L_{v}/L_{h}$ is assumed to be small, $\delta<1$. We define the horizontal Froude number to be $F_{h}=U/NL_{h}$, which is also assumed to be small. Following the above numerical simulations, let $\delta < F_{h}$, which we can also write as $L_{v} < U/N$, i.e. vertical scales are assumed to be smaller than the buoyancy scale. We now define the advective time scale $T=L_{h}/U$. 
To determine the characteristic scale of $W$, we are left with two choices: imposing the scaling from the continuity equation or from the density equation. Previous work \cite{bc2001} chose the latter and obtained a characteristic velocity 
\begin{align}
W \lesssim \frac{RF_{h}g}{\rho_{0}N}\label{scaling1}. 
\end{align}
By contrast, we use the continuity equation (\ref{scaling_cont}), which implies
\begin{align}
W \lesssim \delta U\label{scaling2}.
\end{align}
This scaling for $w$ is consistent with the assumption that $\delta < F_{h}$. Using (\ref{scaling2}), the vertical momentum equation (\ref{scaling_vert}) gives a density scaling of $R\sim \rho_{0}U^{2}/(gL_{v})$. Plugging this result into (\ref{scaling1}), we obtain $W\sim UF_{h}^{2}/\delta$. Because $\delta < F_{h}$ we have $U\delta < UF_{h}^{2}/\delta$ so our assumptions are consistent. Setting $W\sim U\delta$ the horizontal momentum equation (\ref{scaling_horz}) gives $P\sim \rho_{0}U^{2}$. Combining this all, we obtain the following scaling for the Boussinesq equations with $L_{v} < U/N$:  

\begin{align}
\textbf{u}_{h}' = U\textbf{u}_{h},\qquad u_{z}'=U\delta u_{z},\qquad \rho' =\frac{U^{2}\rho_{0}}{gL_{v}}\rho,\qquad p'=\rho_{0}U^{2}p, \nonumber\\
\textbf{x}=L_{h}x,\qquad z'=L_{v}z,\qquad t' = \frac{L_{h}}{U}{t},\qquad Re=\frac{UL_{h}}{\nu},\qquad Sc = \frac{\nu}{D}
\end{align} 
which leads to  
\begin{align}
\frac{\partial \textbf{u}_{h}}{\partial t} + \textbf{u}_{h}\cdot\nabla_{h}\textbf{u}_{h}+u_{z}\frac{\partial \textbf{u}_{h}}{\partial z} &= -\nabla_{h}p + \frac{1}{Re}\nabla_{h}^{2}\textbf{u}_{h} +  \frac{1}{\delta^{2}Re}\frac{\partial^{2}\textbf{u}_{h}}{\partial z^{2}},\\
\delta^{2}\left(\frac{\partial u_{z}}{\partial t} + \textbf{u}_{h}\cdot\nabla_{h}u_{z}+u_{z}\frac{\partial u_{z}}{\partial z}\right) &= -\frac{\partial p}{\partial z} - \rho' + \frac{\delta^{2}}{Re}\nabla_{h}^{2}u_{z} + \frac{1}{Re}\frac{\partial^{2}u_{z}}{\partial z^{2}},\\
\nabla_{h}\cdot\textbf{u}_{h} + \frac{\partial u_{z}}{\partial z} &=1,\\
\frac{\partial \rho'}{\partial t} + \textbf{u}_{h}\cdot\nabla_{h}\rho' + u_{z}\frac{\partial \rho'}{\partial z} -\frac{\delta^{2}}{F_{h}^{2}}u_{z}&=\frac{1}{ReSc}\nabla_{h}^{2}\rho + \frac{1}{\delta^{2}ReSc}\frac{\partial^{2}\rho}{\partial z^{2}},
\end{align}
which holds when $\delta<F_{h}\ll 1$. This suggests that for very small vertical scales with $\delta \ll  F_{h}$ the effects of stratification should be negligible. At such small vertical scales, density variation due to stratification would be negligible and thus we would not expect stratification to play an important role in the overall evolution. Additionally, the presence of the factors of $\delta$ in the denominator of the vertical viscous terms suggests that the effects of viscosity become more dominant at very small vertical scales.  

\begin{figure}
\begin{center}
\includegraphics[scale=0.7]{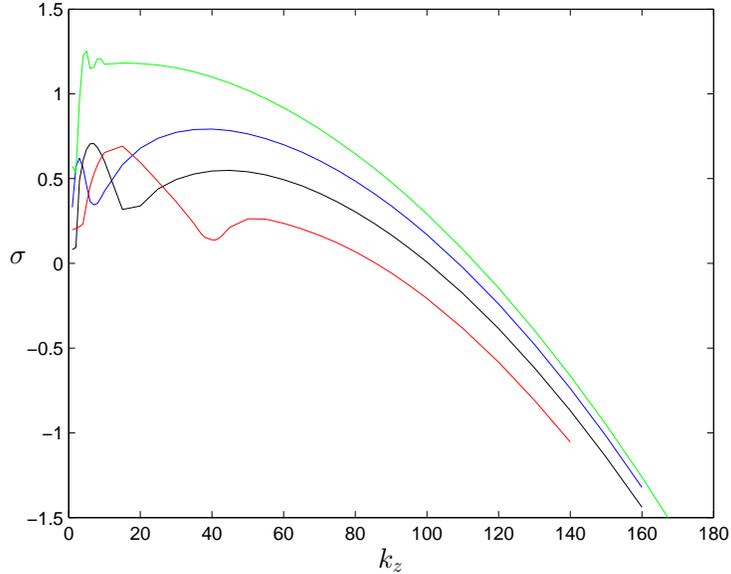}
\caption{Growth rate $\sigma$ as a function $k_{z}$ at $Re=10000$ with $F_{h}=\infty$ (green), $F_{h}=0.2$ (blue), $F_{h}=0.1$ (black), $F_{h}=0.05$ (red)}
\label{sigma_scaling}
\end{center}
\end{figure}

As a result of this scaling analysis we expect that the nature of the instability at short vertical scales to become independent of $F_{h}$ for large $k_{z}$. To test this hypothesis Fig.~\ref{sigma_scaling} shows growth rate as a function of $k_{z}$ for four sets of simulation with $Re=10000$: $F_{h}=0.2,0.1,0.05$ and a new unstratified case with $F_{h}=\infty$ (note that, unlike in Fig.~\ref{FixReVaryFh}, we are not scaling $k_{z}$ by $F_{h}$). The growth rate curves appear to be converging for large $k_{z}$ where $\delta \ll F_{h}$, which agrees with the conclusion of the above scaling analysis. These large $k_{z}$ are well into the viscous damping range and as discussed above, the effects of viscosity become stronger and we observe a sharper decrease in the growth rate.

For the short-wave instability examined above, $\delta/F_{h}=1/(k_{z}F_{h})$ ranges from $\approx 0.5$ down to $0.1$, which is $<1$ but not $\ll 1$. As a result, we do not necessarily expect the characteristics of this instability to be independent of $F_{h}$ for the parameters considered here. Indeed, our stability analysis shows that the (unscaled) wavenumber $k_{z}$ of the short-wave peak is weakly dependent on $F_{h}$, through the $F_{h}^{1/5}$ factor in (\ref{buoyscale}). However, by examining even larger $k_{z}F_{h}$ (i.e. even smaller $\delta/F_{h}$), this scale analysis suggests that the nature of the short-wave instability will eventually become independent of $F_{h}$.  

%%%%%%%%%%%%%%%%%%%%%%%%%%%%%%
% CONCLUSION
%%%%%%%%%%%%%%%%%%%%%%%%%%%%%%

\section{Conclusions}
In this paper, we have investigated the linear stability of the Lamb-Chapylgin dipole for perturbations with small vertical scales. In particular , we have considered vertical scales from around the buoyancy scale $U/N$, where the zigzag instability occurs \cite{bc2000c,bc2001,waitebartello2004,waite2011}, down to the dissipation scale. We have discovered a short-wave instability that emerges at scales much smaller than the buoyancy scale. This instability exhibits a growth rate that is comparable to, and possibly even greater than, that of the zigzag instability. Despite having a similar growth rate in some cases, the structure of the instability is qualitatively different that of the zigzag peak suggesting a different mechanism is governing the evolution.  We have discovered that the location of the peak depends upon a combination of the Reynolds and Froude numbers, specifically the buoyancy Reynolds number $Re_{b}$ which plays an important role in stratified fluids. The wavenumber of maximum growth rate for the short-wave instability is found to scale like $F_{h}k_{z}\sim Re_{b}^{2/5}$ for the range of $Re_{b}$ considered here. We expect this may change at even larger $Re_{b}$. By contrast, the maximum growth rate of the zigzag instability occurs for $F_{h}k_{z}\sim 1$ \cite{bc2000b}. As a result, these instabilities will be widely separated when $Re_{b}\gg 1$, as in the case of strongly stratified turbulence \cite{brethouwer2007}.

This new instability has implications for numerical modelling of small scales in stratified turbulence as it provides an additional mechanism for the transfer of energy to small vertical scales. In nature, perturbations are broad-band and so short vertical scales will be excited. Our results show that such short scales may grow, at least initially, as fast as the zigzag instability. Important questions to be addressed in future work are how does this short-wave instability evolve nonlinearly, and how does it saturate? There is some suggestion that such perturbations may saturate at a relatively low level \cite{nganstraubbartello2005,waitesmol2008} but this question requires further study. 

\section*{Acknowledgments}
Financial support for this work was provided by the Natural Sciences and Engineering Research Council of Canada. Computations were made possible by the facilities of the Shared Hierarchical Academic Research Computing Network (SHARCNET) and Compute/Calcul Canada.

%%%%%%%%%%%%%%%%%%%%%%%%%%%%%%
% BIBLIOGRAPHY
%%%%%%%%%%%%%%%%%%%%%%%%%%%%%%

\bibliographystyle{elsarticle-num}
\bibliography{research}

\end{document}